\documentclass[twocolumn, aps, prb]{revtex4}
\usepackage{graphicx}
\usepackage{amssymb}
\usepackage{amsmath}

\def\Journal #1,#2,#3,#4#5#6#7{#1 {\bf #2}, #3 (#4#5#6#7)}
\def\lsim{\lower -0.3ex \hbox{$<$} \kern -0.75em \lower 0.7ex \hbox{$\sim$}}
\def\gsim{\lower -0.3ex \hbox{$>$} \kern -0.75em \lower 0.7ex \hbox{$\sim$}}

\newcommand{\GVec}[1]{\mbox{\boldmath$#1$}}

\def\Vec#1{{\bf #1}}
\def\GVec#1{\mbox{\boldmath $#1$}}

\def\vare{\varepsilon}

\begin{document}

\title{
% Landau level structure in monolayer-bilayer junction
%Electronic structure of monolayer-bilayer junction in magnetic fields
Interface Landau levels in graphene monolayer-bilayer junction
}
\author{Mikito Koshino$^1$, Takeshi Nakanishi$^2$, and Tsuneya Ando$^1$}
\affiliation{
$^1$Department of Physics, Tokyo Institute of Technology\\
2-12-1 Ookayama, Meguro-ku, Tokyo 152-8551, Japan\\
$^2$Nanotube Research Center, AIST, 1--1--1 Higashi, Tsukuba 305-8565, Japan}

\date{\today}

%%%%%%%%%%%%%%%%%%%%%%%%%%%%%%%%%%%%%%%%%%%%%%%%%%%%%%%%%%%%%%%%%%%%%%%%%%%%%%%
%
%%%%%%%%%%%%%%%%%%%%%%%%%%%%%%%%%%%%%%%%%%%%%%%%%%%%%%%%%%%%%%%%%%%%%%%%%%%%%%%
\begin{abstract}
Electronic structure of graphene monolayer-bilayer junction
in a magnetic field is studied within an effective-mass approximation.
The energy spectrum is characterized 
by interface Landau levels, i.e., the locally flat bands
appearing near the boundary region,
resulting in a series of characteristic peaks in the local density of states.
Their energies are independent of boundary types such as zigzag or armchair.
In the atomic scale, the local density of states shows a Kekul\'{e} pattern
due to the valley mixing in the armchair boundary, while does not in the zigzag boundary.
\end{abstract}
%%%%%%%%%%%%%%%%%%%%%%%%%%%%%%%%%%%%%%%%%%%%%%%%%%%%%%%%%%%%%%%%%%%%%%%%%%%%%%%
%
%%%%%%%%%%%%%%%%%%%%%%%%%%%%%%%%%%%%%%%%%%%%%%%%%%%%%%%%%%%%%%%%%%%%%%%%%%%%%%%
\maketitle
%%%%%%%%%%%%%%%%%%%%%%%%%%%%%%%%%%%%%%%%%%%%%%%%%%%%%%%%%%%%%%%%%%%%%%%%%%%%%%%
%
%%%%%%%%%%%%%%%%%%%%%%%%%%%%%%%%%%%%%%%%%%%%%%%%%%%%%%%%%%%%%%%%%%%%%%%%%%%%%%%
\section{Introduction}
\label{sec_intro}
%%%%%%%%%%%%%%%%%%%%%%%%%%%%%%%%%%%%%%%%%%%%%%%%%%%%%%%%%%%%%%%%%%%%%%%%%%%%%%%
%
%%%%%%%%%%%%%%%%%%%%%%%%%%%%%%%%%%%%%%%%%%%%%%%%%%%%%%%%%%%%%%%%%%%%%%%%%%%%%%%
Graphene\cite{McClure_1956a,Slonczewski_and_Weiss_1958a,DiVincenzo_and_Mele_1984a,Shon_and_Ando_1998a,Zheng_and_Ando_2002a,Gusynin_and_Sharapov_2005a,Peres_et_al_2006a,Ando_2005a} and its bilayer\cite{McCann_and_Falko_2006a,Koshino_and_Ando_2006a,Katsnelson_2006b,McCann_2006a,Nilsson_et_al_2006b,Guinea_et_al_2006a,Castro_et_al_2007a,Koshino_2009a} are characterized by zero-gap band structures supporting different types of chiral particles and Landau-level structures.
%In magnetic fields, they host unusual Landau-level structures distinguished from each other and exhibit the integer quantum Hall effect with peculiar plateau structures.\cite{McClure_1956a,Shon_and_Ando_1998a,Zheng_and_Ando_2002a,Gusynin_and_Sharapov_2005a,Peres_et_al_2006a,McCann_and_Falko_2006a,Guinea_et_al_2006a}
Recently, atomically thin graphene samples were experimentally fabricated using mechanical exfoliation\cite{Novoselov_et_al_2004a,Novoselov_et_al_2006a} and epitaxial growth.\cite{Berger_et_al_2004a,Ohta_et_al_2006a}
The characteristic Landau-level structure and integer quantum Hall effect\cite{McClure_1956a,Shon_and_Ando_1998a,Zheng_and_Ando_2002a,Gusynin_and_Sharapov_2005a,Peres_et_al_2006a,McCann_and_Falko_2006a,Guinea_et_al_2006a} were observed in magnetotransport measurements.\cite{Novoselov_et_al_2005a,Zhang_et_al_2005a}
In this paper we study the electronic structure of a hybrid system composed of monolayer and bilayer graphenes in magnetic fields.
\par
%%%%%%%%%%%%%%%%%%%%%%%%%%%%%%%%%%%%%%%%%%%%%%%%%%%%%%%%%%%%%%%%%%%%%%%%%%%%%%%
%
%%%%%%%%%%%%%%%%%%%%%%%%%%%%%%%%%%%%%%%%%%%%%%%%%%%%%%%%%%%%%%%%%%%%%%%%%%%%%%%
The band structure of monolayer graphene is characterized by 
Dirac-like spectrum in which conduction and valence bands 
with linear dispersion stick 
at the $K$ and $K'$ points located at a Brillouin zone corner,
\cite{McClure_1956a,Slonczewski_and_Weiss_1958a,
DiVincenzo_and_Mele_1984a,Ando_2005a} which are called valleys.
Bilayer graphene has a zero-gap structure, but 
with quadratic dispersion unlike monolayer.
\cite{McCann_and_Falko_2006a,Koshino_and_Ando_2006a,
Katsnelson_2006b,McCann_2006a,Nilsson_et_al_2006b,Guinea_et_al_2006a,Castro_et_al_2007a,Koshino_2009a}
In a magnetic field, the level structure of monolayer 
\cite{McClure_1956a,Shon_and_Ando_1998a,Zheng_and_Ando_2002a,
Gusynin_and_Sharapov_2005a,Peres_et_al_2006a}
and bilayer \cite{McCann_and_Falko_2006a,Castro_et_al_2007a,
Koshino_and_Ando_2008a,Koshino_and_McCann_2010a}
differs in number of degeneracy at zero-energy
and the quantum Hall plateaus appear at different filling factors
accordingly. \cite{Novoselov_et_al_2005a,Zhang_et_al_2005a,Castro_et_al_2007a}
\par
%%%%%%%%%%%%%%%%%%%%%%%%%%%%%%%%%%%%%%%%%%%%%%%%%%%%%%%%%%%%%%%%%%%%%%%%%%%%%%%
%
%%%%%%%%%%%%%%%%%%%%%%%%%%%%%%%%%%%%%%%%%%%%%%%%%%%%%%%%%%%%%%%%%%%%%%%%%%%%%%%
The electronic states of graphene with an edge have been 
studied in theories.\cite{Fujita_et_al_1996a,Nakada_et_al_1996a,Wakabayashi_et_al_1999a,Wakabayashi_and_Sigrist_2000a,Wakabayashi_2001a,Wakabayashi_2002a,McCann_and_Falko_2004a,Brey_and_Fertig_2006b,Peres_et_al_2006b,Son_et_al_2006b,Son_et_al_2006c,Obradovic_et_al_2006a,Yang_et_al_2007a,Yang_et_al_2008b,Li_and_Lu_2008a,Raza_and_Kan_2008a,Ryzhii_et_al_2008a,Wassmann_et_al_2008a,Akhmerov_and_Beenakker_2008a,Nguyen_et_al_2009a,Gunlycke_and_White_2010a}
In particular, when the boundary is along zigzag direction,
special states localized at the edge appear
as zero-energy modes.
\cite{Fujita_et_al_1996a,Nakada_et_al_1996a}
Similar zero-energy edge states exist also in
bilayer graphene. \cite{Castro_et_al_2008a,Sahu_et_al_2008a}
In a magnetic field, electron-like and hole-like
Landau levels are shifted 
upwards and downwards near the boundary, respectively, 
forming edge channels away from zero energy.
\cite{Brey_and_Fertig_2006b,Abanin_et_al_2006a,Abanin_et_al_2007a,Abanin_et_al_2007b,Castro_et_al_2007a}
Recently, the transport through 
quantum structures consisting of monolayer and bilayer graphenes 
was investigated. \cite{Nilsson_et_al_2007a,Gonzalez_et_al_2010a}
In a previous paper, the boundary condition between monolayer and
bilayer graphenes connected by a monoatomic step 
was studied, and the transmission probability through the
junction was calculated in the absence of magnetic field. 
\cite{Nakanishi_et_al_2010a}
\par
%%%%%%%%%%%%%%%%%%%%%%%%%%%%%%%%%%%%%%%%%%%%%%%%%%%%%%%%%%%%%%%%%%%%%%%%%%%%%%%
%
%%%%%%%%%%%%%%%%%%%%%%%%%%%%%%%%%%%%%%%%%%%%%%%%%%%%%%%%%%%%%%%%%%%%%%%%%%%%%%%
In this paper, we study the energy spectrum and local density of states
of the monolayer-bilayer graphene junction in magnetic fields.
Based on the previous study,\cite{Nakanishi_et_al_2010a} 
we consider a composed system of half-infinite
graphene monolayer and bilayer
connected by a mono-atomic step along zigzag or armchair direction.
In Sec.\ \ref{sec_eff}, we present effective mass description for 
monolayer and bilayer graphenes and introduce formulation to
describe Landau levels of the junction in Sec.\ \ref{sec_junc}.
In Sec.\ \ref{sec_num},
we numerically calculate 
the energy spectra for several types of the boundaries
as well as the local density of states.
The conclusion is presented in Sec.\ \ref{sec_conc}.
\par
%%%%%%%%%%%%%%%%%%%%%%%%%%%%%%%%%%%%%%%%%%%%%%%%%%%%%%%%%%%%%%%%%%%%%%%%%%%%%%%
%
%%%%%%%%%%%%%%%%%%%%%%%%%%%%%%%%%%%%%%%%%%%%%%%%%%%%%%%%%%%%%%%%%%%%%%%%%%%%%%%
\section{Effective mass Hamiltonian}
\label{sec_eff}
\subsection{Monolayer graphene}
%%%%%%%%%%%%%%%%%%%%%%%%%%%%%%%%%%%%%%%%%%%%%%%%%%%%%%%%%%%%%%%%%%%%%%%%%%%%%%%
%
%%%%%%%%%%%%%%%%%%%%%%%%%%%%%%%%%%%%%%%%%%%%%%%%%%%%%%%%%%%%%%%%%%%%%%%%%%%%%%%
Graphene is composed of a honeycomb network of carbon atoms, containing a pair of sublattices, denoted by $A$ and $B$.
Electronic states in the vicinity of $K$ and $K'$ points in the Brillouin zone are well described by envelope functions $(F_A^K,F^K_B)$ and $(F^{K'}_{A},F^{K'}_{B})$, respectively, in an effective-mass approximation.
At the K point, the effective Hamiltonian for
$(F_A^K,F^K_B)$ is \cite{McClure_1956a,Slonczewski_and_Weiss_1958a,DiVincenzo_and_Mele_1984a,Ando_2005a}
\begin{eqnarray}
{\mathcal H}^K = 
\begin{pmatrix} 0 & v \pi_- \\ v \pi_+ & 0 
\end{pmatrix} ,
\label{eq_H_mono}
\end{eqnarray}
where $v \approx 1\times 10^6$ m/s is the band velocity
%\cite{Novoselov_et_al_2005a,Zhang_et_al_2005a}, 
$\pi_\pm = \pi_x \pm i \pi_y$, and $\GVec{\pi} = -i\hbar \GVec{\nabla} +
(e/c) {\bf A}$ with vector potential in the Landau gauge, 
${\bf A} = (0,Bx)$, 
giving external magnetic field ${\bf B} = \GVec{\nabla}\times {\bf A}$.
The Hamiltonian at the $K'$ point is obtained by exchanging $\pi_\pm$ in
Eq.\ (\ref{eq_H_mono}).
\par
%%%%%%%%%%%%%%%%%%%%%%%%%%%%%%%%%%%%%%%%%%%%%%%%%%%%%%%%%%%%%%%%%%%%%%%%%%%%%%%
%
%%%%%%%%%%%%%%%%%%%%%%%%%%%%%%%%%%%%%%%%%%%%%%%%%%%%%%%%%%%%%%%%%%%%%%%%%%%%%%%
The wavenumber $k_y$ remains a good quantum number in the present geometry.
The operator $\pi_\pm$ can be expressed as
\begin{equation}
\begin{array}{l}
v \pi_+ = i \hbar\omega_B a^\dagger, \\
v \pi_- = -i \hbar\omega_B a,
\end{array}
\end{equation}
where $\hbar\omega_B = \sqrt{2}\hbar v/l_B$
with magnetic length $l_{B}=\sqrt{c\hbar /(eB)}$ and $a^\dagger$ and $a$ are raising and lowering operators, respectively, defined by
\begin{equation}
 a = \frac{\partial}{\partial z} + \frac{z}{2},
\end{equation}
with dimensionless coordinate,
\begin{equation}
z = \sqrt{2} \left(\frac{x}{l_B} + k_y l_B\right) = \frac{\sqrt{2}(x-X)}{l_B} .
\end{equation}
Here, the center coordinate of the cyclotron motion is defined by
\begin{equation}
X = -k_yl_B^2 .
\end{equation}

The Schr\"{o}dinger equation then becomes
\begin{equation}
\begin{array}{l}
\varepsilon F^K_A = -i \hbar\omega_B a F^K_B \\
\noalign{\vspace{0.10cm}}
\varepsilon F^K_B = i \hbar\omega_B a^\dagger F^K_A,
\end{array}
\label{eq_schr}
\end{equation}
giving
\begin{eqnarray}
(\nu -a^\dagger a) F^K_B = 
\left( 
\frac{\partial^2 }{\partial z^2} 
+ \nu + \frac 1 2 - \frac{z^2}{4}\right) F^K_B = 0,
\label{eq_weber_diff}
\end{eqnarray}
with 
\begin{equation}
 \nu = \Big( {\varepsilon \over \hbar\omega_B } \Big)^2.
\label{eq_nu_mono}
\end{equation}
The independent solutions of Eq.\ (\ref{eq_weber_diff}) 
are given by $D_\nu(z)$ and $D_{-\nu-1}(-iz)$,
where $D_\nu(z)$ is Weber's parabolic cylinder function
defined by
\begin{eqnarray}
D_\nu(z) &=& 
2^{\nu/2}\sqrt{\pi}e^{-z^2/4}
\Biggl[
\frac{1}{\Gamma((1-\nu)/2)}
F\left(-\frac{\nu}{2},\frac 1 2 ; \frac{z^2}{2}\right)
\nonumber\\
&& 
-\frac{\sqrt{2}z}{\Gamma(-\nu/2)}
F\left(\frac{1-\nu}{2},\frac 3 2 ; \frac{z^2}{2}\right)
\Biggr],
\label{eq_weber}
\end{eqnarray}
with $F(\alpha,\gamma;z)$ being Kummer's hypergeometric function.
The components $F_K^A$ and $F_K^B$ are related
by Eq.\ (\ref{eq_schr}) with formula
\begin{equation}
\begin{array}{rcl}
a^\dagger D_\nu(z) & = & D_{\nu+1}(z), \\
a D_\nu(z) & = & \nu D_{\nu-1}(z).
\end{array}
\label{eq_deriv}
\end{equation}

Because of relation
\begin{equation}
 D_{\nu}(-z) = e^{\nu\pi i}D_{\nu}(z)
+ 
\frac{\sqrt{2\pi}}{\Gamma(-\nu)}
e^{(\nu+1)\pi i/2}D_{-\nu-1}(-iz),
\end{equation}
$D_\nu(z)$ and $D_{\nu}(-z)$ can also be chosen as
independent solutions of Eq.\ (\ref{eq_weber_diff}),
as long as $1/\Gamma(-\nu)$ is nonzero, i.e.,
$\nu$ is not 0 or a positive integer.
Then, $D_\nu(z)$ and $D_\nu(-z)$ exponentially diverge
in limits $z = -\infty$ and $+\infty$, respectively,
while converge to zero in the opposite side.
They can never be a bulk eigenfunction,
but may appear when the system is half-infinite in the $x$ direction.
For a nonnegative integer $n$, 
$D_n(z)$ and $D_n(-z)$ are linearly dependent, and 
coincide with usual Landau-level function
except for a normalization factor as
\begin{equation}
 D_n(z) = (-1)^n D_n(-z) = 2^{-n/2} e^{-z^2/4} H_n(z/\sqrt{2}),
\end{equation}
with Hermite polynomial $H_n(z)$.
The other solution $D_{-n-1}(i z)$ then diverges 
both in limits $z = +\infty$ and $-\infty$ and is excluded.
$D_n(z)$ at a negative integer $n$
generally diverges for $z \to -\infty$.
At $n=-1$, for example, we have
\begin{equation}
 D_{-1}(z) = \sqrt{2\pi}\,e^{z^2/4} 
\left[
-1+\textrm{erf}(z/\sqrt{2})
\right],
\label{eq_D_-1}
\end{equation}
with error function
\begin{equation}
{\rm erf}(x) = \int_0^x e^{t^2} d t .
\end{equation}
\par
%%%%%%%%%%%%%%%%%%%%%%%%%%%%%%%%%%%%%%%%%%%%%%%%%%%%%%%%%%%%%%%%%%%%%%%%%%%%%%%
%
%%%%%%%%%%%%%%%%%%%%%%%%%%%%%%%%%%%%%%%%%%%%%%%%%%%%%%%%%%%%%%%%%%%%%%%%%%%%%%%
Let us define 
\begin{equation}
\begin{array}{l}
\phi^R_{\nu}(z) = D_\nu (z), \\
\phi^L_{\nu}(z) = D_\nu (-z),
\end{array}
\end{equation}
where $L$ and $R$ represent the solutions finite in limits $z \to -\infty$ and $+\infty$, respectively.
We will consider a monolayer-bilayer junction
in which the region $x<0$ is monolayer and $x>0$ is bilayer.
The eigen function in monolayer is given by
\begin{eqnarray}
&& \left(
\begin{array}{c}
F_A^K
\\
F_B^K
\end{array}
\right)
=
\left(
\begin{array}{c}
 i \alpha_1 \phi^L_{\nu-1} 
\\
\alpha_2 \phi^L_{\nu} 
\end{array}
\right) e^{-i X y/ l_B^2 } ,
\\
&& \left(
\begin{array}{c}
\alpha_1
\\
\alpha_2
\end{array}
\right)
=
\left(
\begin{array}{c}
\varepsilon/\hbar\omega_B
\\
1
\end{array}
\right).
\end{eqnarray}
The wavefunction at the $K'$ point can be obtained by
$(F_A^{K'},F_B^{K'})=(F_B^{K},F_A^{K})$.
\par
\begin{figure}
\includegraphics[width=70mm]{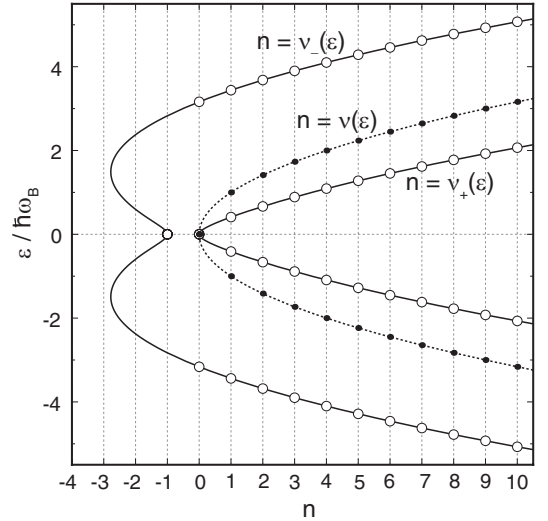}
\caption{Plots of $\nu(\varepsilon)$,
$\nu_+(\varepsilon)$, and $\nu_-(\varepsilon)$
with energy $\varepsilon$ set to the vertical axis. 
Black and white circles represent the 
Landau levels of bulk monolayer and bilayer, respectively.
$\gamma_1/\hbar\omega_B = 3$ is taken for bilayer.
}
\label{fig_nu}
\end{figure}

%%%%%%%%%%%%%%%%%%%%%%%%%%%%%%%%%%%%%%%%%%%%%%%%%%%%%%%%%%%%%%%%%%%%%%%%%%%%%%%
%
%%%%%%%%%%%%%%%%%%%%%%%%%%%%%%%%%%%%%%%%%%%%%%%%%%%%%%%%%%%%%%%%%%%%%%%%%%%%%%%
The Landau level energies of bulk monolayer graphene are given by the 
condition that the wavefunction 
is finite in limits $x=\pm\infty$, i.e., $\nu(\varepsilon)$ 
is non-negative integer $n$. 
We get \cite{McClure_1956a,Shon_and_Ando_1998a,Zheng_and_Ando_2002a,
Gusynin_and_Sharapov_2005a,Peres_et_al_2006a}
\begin{equation}
\begin{array}{l}
\varepsilon_0 = 0, \\
\varepsilon_{n,\pm} = \pm \hbar\omega_B \sqrt{n} \quad (n=1,2,\cdots).
\end{array}
\end{equation}
The plot of $\nu(\varepsilon)$ and bulk Landau-level energies are 
shown in Fig.\ \ref{fig_nu}.
\par
%%%%%%%%%%%%%%%%%%%%%%%%%%%%%%%%%%%%%%%%%%%%%%%%%%%%%%%%%%%%%%%%%%%%%%%%%%%%%%%
%
%%%%%%%%%%%%%%%%%%%%%%%%%%%%%%%%%%%%%%%%%%%%%%%%%%%%%%%%%%%%%%%%%%%%%%%%%%%%%%%
%%%%%%%%%%%%%%%%%%%%%%%%%%%%%%%%%%%%%%%%%%%%%%%%%%%%%%%%%%%%%%%%%%%%%%%%%%%%%%%
%
%%%%%%%%%%%%%%%%%%%%%%%%%%%%%%%%%%%%%%%%%%%%%%%%%%%%%%%%%%%%%%%%%%%%%%%%%%%%%%%
\subsection{Bilayer graphene}
\label{sec_bi}
%%%%%%%%%%%%%%%%%%%%%%%%%%%%%%%%%%%%%%%%%%%%%%%%%%%%%%%%%%%%%%%%%%%%%%%%%%%%%%%
%
%%%%%%%%%%%%%%%%%%%%%%%%%%%%%%%%%%%%%%%%%%%%%%%%%%%%%%%%%%%%%%%%%%%%%%%%%%%%%%%
Bilayer graphene is a pair of graphene layers arranged in AB (Bernal)
stacking and includes $A_1$ and $B_1$ atoms on layer 1 and $A_2$ and
$B_2$ on layer 2.
The layers are arranged such that sites
$B_1$ and $A_2$ are directly below and above each other, which are
connected by interlayer coupling $\gamma_1 \sim 0.39$ eV. \cite{Misu_et_al_1979a}
The effective Hamiltonian for $(F_{A1}^K,F^K_{B1},F_{A2}^K,F^K_{B2})$ is given by \cite{McCann_and_Falko_2006a,Koshino_and_Ando_2006a,Katsnelson_2006b,McCann_2006a,Nilsson_et_al_2006b,Guinea_et_al_2006a,Castro_et_al_2007a,Koshino_2009a}
%%%%%%%%%%%%%%%%%%%%%%%%%%%%%%%%%%%%%%%%%%%%%%%%%%%%%%%%%%%%%%%%%%%%%%%%%%%%%%%
%
%%%%%%%%%%%%%%%%%%%%%%%%%%%%%%%%%%%%%%%%%%%%%%%%%%%%%%%%%%%%%%%%%%%%%%%%%%%%%%%
\begin{eqnarray}
{\cal H}^K =
\begin{pmatrix} 0 & v \pi_- & 0 & 0 \\
v \pi_+  & 0 & \gamma_1 & 0 \\
0  & \gamma_1 & 0 &  v \pi_- \\
0 & 0 &  v \pi_+ & 0
\end{pmatrix} .
\label{eq_H_bi}
\end{eqnarray}
%%%%%%%%%%%%%%%%%%%%%%%%%%%%%%%%%%%%%%%%%%%%%%%%%%%%%%%%%%%%%%%%%%%%%%%%%%%%%%%
%
%%%%%%%%%%%%%%%%%%%%%%%%%%%%%%%%%%%%%%%%%%%%%%%%%%%%%%%%%%%%%%%%%%%%%%%%%%%%%%%
The Hamiltonian at the $K'$ point is obtained by exchanging $\pi_\pm$ in Eq.\ (\ref{eq_H_bi}). 
\par
%%%%%%%%%%%%%%%%%%%%%%%%%%%%%%%%%%%%%%%%%%%%%%%%%%%%%%%%%%%%%%%%%%%%%%%%%%%%%%%
%
%%%%%%%%%%%%%%%%%%%%%%%%%%%%%%%%%%%%%%%%%%%%%%%%%%%%%%%%%%%%%%%%%%%%%%%%%%%%%%%
The eigenfunction of Eq.\ (\ref{eq_H_bi}) 
finite in limit $x \to +\infty$ is written as
\begin{equation}
\left(
\begin{array}{c}
F_{A1}^K
\\
F_{B1}^K
\\
F_{A2}^K
\\
F_{B2}^K
\end{array}
\right)
=
\left(
\begin{array}{c}
- i \beta^\mu_1 \,\phi^R_{\nu_\mu-1} 
\\
\beta^\mu_2 \,\phi^R_{\nu_\mu} 
\\
 \beta^\mu_3 \,\phi^R_{\nu_\mu} 
\\
i \beta^\mu_4 \,\phi^R_{\nu_\mu+1} 
\end{array}
\right) e^{ -i X y / l_B^2} ,
\label{eq_wave_bi}
\end{equation}
with
\begin{eqnarray}
&&\nu_\mu(\varepsilon) = 
-\frac{1}{2} + 
\tilde\varepsilon^2
+ \frac{\mu}{2}
\sqrt{4\tilde\varepsilon^2 \tilde\gamma_1^2 + 1},
\label{eq_nu_pm}
\\
&&\left(
\begin{array}{c}
\beta^\mu_1
\\
\beta^\mu_2
\\
\beta^\mu_3
\\
\beta^\mu_4
\end{array}
\right)
=
\left(
\begin{array}{c}
\tilde\gamma_1 \nu_\mu \tilde\varepsilon/(\tilde\varepsilon^2 - \nu_\mu)
\\
 \tilde\gamma_1 \tilde\varepsilon^2/(\tilde\varepsilon^2 - \nu_\mu)
\\
 \tilde\varepsilon 
\\
1
\end{array}
\right),
\label{eq_beta}
\end{eqnarray}
where $\mu=\pm$ is another degree of freedom,
$\tilde\varepsilon = \varepsilon/(\hbar\omega_B)$, and
$\tilde\gamma_1 = \gamma_1/(\hbar\omega_B)$.
The wavefunction at the $K'$ point is obtained by
$(F_{A1}^{K'},F_{B1}^{K'},F_{A2}^{K'},F_{B2}^{K'})=(F_{B2}^{K},F_{A2}^{K},F_{B1}^{K},F_{A1}^{K})$.

The Landau levels of bulk bilayer graphene are obtained by the 
condition that the wavefunction of Eq.\ (\ref{eq_wave_bi})
is finite in limits $x \to \pm\infty$, i.e.,
includes only $\phi^{R}_{n} $ of non-negative integer $n$.
Allowed indexes are $\nu_+(\varepsilon) = 0,1,2,\cdots$ and
$\nu_-(\varepsilon) = -1,0,1,\cdots$.
For $\nu_-=-1,0$ and $\nu_+ =0$, the wavefunction Eq.\ (\ref{eq_wave_bi}) 
appears to include $\phi^{R}_{n}$ with negative $n$,
but corresponding coefficient such as $\beta^\pm_1$ vanishes.
At $\varepsilon=0$, there are two energy levels for $\nu_+ = 0$ and $\nu_- = -1$. \cite{McCann_and_Falko_2006a}
Figure \ref{fig_nu} illustrates $\nu_\pm(\varepsilon)$
and bulk Landau-level energies of bilayer graphene.

\section{Monolayer-bilayer junction}
\label{sec_junc}

\begin{figure}
\includegraphics[width=60mm]{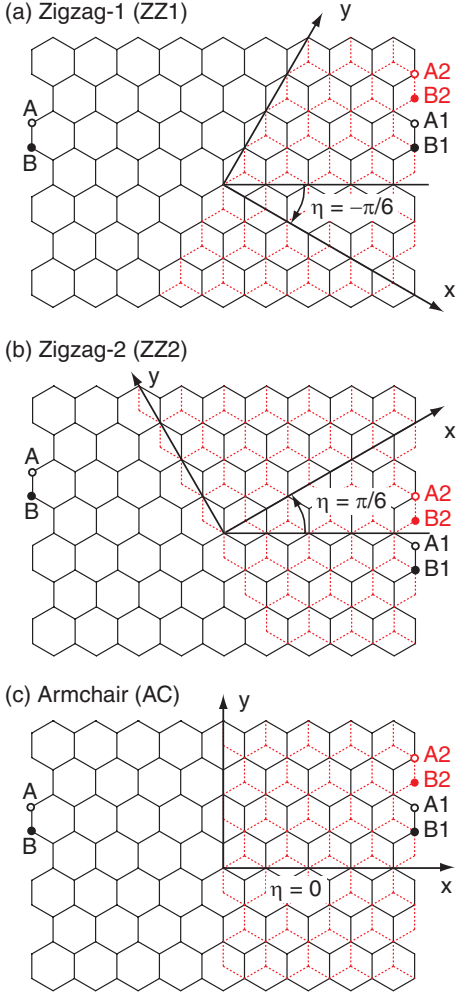}
\caption{
Monolayer-bilayer graphene junctions
with boundary types of (a) ZZ1, (b) ZZ2, and (c) AC.
}
\label{fig_atom}
\end{figure}

We consider a composite system of monolayer and 
bilayer graphenes, where the left half ($x<0$)
is monolayer and the right half ($x>0$) is AB-stacked bilayer.
We assume that one layer of the bilayer part, 
containing $A_1$ and $B_1$ sites,
seamlessly continues to the monolayer part
with $A$ and $B$ sites,
while the other layer composed of $A_2$ and $B_2$ sites
is sharply cut at the boundary chosen as $x=0$.
In the following, we consider two kinds of zigzag boundaries,
zigzag-1 (ZZ1) and zigzag-2 (ZZ2)
and armchair boundary (AC) as illustrated in 
Fig.\ \ref{fig_atom} (a), (b), and (c).\cite{Nakanishi_et_al_2010a}

We assume that the system infinitely continues in the $y$ direction
parallel to the boundary.
The wavefunctions of monolayer and bilayer regions
are required to converge in limits $x=-\infty$ and $\infty$,
respectively.
At given energy $\varepsilon$,
they can be written for the monolayer part $(x<0)$
%%%%%%%%%%%%%
\begin{eqnarray}
&& \left(
\begin{array}{c}
F_A^K
\\
F_B^K
\end{array}
\right)
=
A^K \left(
\begin{array}{c}
 i \alpha_1 \phi^L_{\nu-1} ,
\\
\alpha_2 \phi^L_{\nu} 
\end{array}
\right) e^{ -i X y / l_B^2 } ,
\label{eq_wave_mono_K}
\\
&& \left(
\begin{array}{c}
F_A^{K'}
\\
F_B^{K'}
\end{array}
\right)
=
A^{K'} \left(
\begin{array}{c}
-i \alpha_2 \phi^L_{\nu} 
\\
\alpha_1 \phi^L_{\nu-1} 
\end{array}
\right)  e^{ -i X y / l_B^2 } ,
\label{eq_wave_mono_K'}
\end{eqnarray}
%%%%%%%%%%%%%
and for the bilayer part $(x>0)$
%%%%%%%%%%%%%
\begin{eqnarray}
&& \left(
\begin{array}{c}
F_{A1}^K
\\
F_{B1}^K
\\
F_{A2}^K
\\
F_{B2}^K
\end{array}
\right)
=
\sum_{\mu=\pm}
B^K_\mu
\left(
\begin{array}{c}
- i \beta^\mu_1 \,\phi^R_{\nu_\mu-1}
\\
\beta^\mu_2 \,\phi^R_{\nu_\mu} 
\\
 \beta^\mu_3 \,\phi^R_{\nu_\mu} 
\\
i \beta^\mu_4 \,\phi^R_{\nu_\mu+1} 
\end{array}
\right) e^{ -i X y / l_B^2 } ,
\label{eq_wave_bi_K}
\\
&& \left(
\begin{array}{c}
F_{A1}^{K'}
\\
F_{B1}^{K'}
\\
F_{A2}^{K'}
\\
F_{B2}^{K'}
\end{array}
\right)
=
\sum_{\mu=\pm}
B^{K'}_\mu
\left(
\begin{array}{c}
 i \beta^\mu_4 \,\phi^R_{\nu_\mu+1} 
\\
\beta^\mu_3 \,\phi^R_{\nu_\mu} 
\\
 \beta^\mu_2 \,\phi^R_{\nu_\mu} 
\\
-i \beta^\mu_1 \,\phi^R_{\nu_\mu-1} 
\end{array}
\right) e^{ -i X y / l_B^2 } , \qquad
\label{eq_wave_bi_K'}
\end{eqnarray}
%%%%%%%%%%%%%
with six unknown coefficients
$A^K$, $A^{K'}$, $B^K_\pm$, and $B^{K'}_\pm$ 
to be determined by the specific boundary condition.

\subsection{Zigzag boundary, ZZ1}

The boundary ZZ1 is parallel to 
the zigzag direction of honeycomb lattice,
and the front-most line of bilayer part
is formed by $B_1$ and $A_2$ sites.
As the zigzag boundary does not mix the wavefunctions
at different valleys $K$ and $K'$,
the boundary condition is separately expressed for each valley.
The conditions are\cite{Nakanishi_et_al_2010a}
\begin{eqnarray}
&& F^v_{A1}(0,y) =  F^v_A(0,y), \nonumber\\
&& F^v_{B1}(0,y) =  F^v_B(0,y), \\
&& F^v_{B2}(0,y) =  0, \nonumber
\label{eq_bc_zz1}
\end{eqnarray}
for $v=K$ and $K'$.
For the $K$ point, the conditions are rewritten 
with use of the wavefunctions 
(\ref{eq_wave_mono_K}) and (\ref{eq_wave_bi_K}) as
\begin{eqnarray}
M^K_{\rm ZZ1}\left(
\begin{array}{c}
A^K
\\
B^K_+
\\
B^K_-
\end{array}
\right)
= 0,
\end{eqnarray}
with
\begin{equation}
M^K_{\rm ZZ1} \equiv
\left(
\begin{array}{ccc}
\alpha_1 \phi^L_{\nu-1} & 
\beta^+_1 \,\phi^R_{\nu_+ -1} &
\beta^-_1 \,\phi^R_{\nu_- -1} 
\\
- \alpha_2 \phi^L_{\nu} & 
\beta^+_2 \,\phi^R_{\nu_+} &
\beta^-_2 \,\phi^R_{\nu_-} 
\\
0 & 
\beta^+_4 \,\phi^R_{\nu_+ +1} &
\beta^-_4 \,\phi^R_{\nu_- +1} 
\end{array}
\right),
\label{eq_eigen_zz1_K}
\end{equation}
where the wavefunctions such as
$\phi^R_{\nu_\mu}$ represent the values at $x=0$.
For each $X$, the eigen energies are obtained by
searching for solutions of $\det M^K_{\rm ZZ1} = 0$.
The corresponding equation for $K'$ is
\begin{eqnarray}
& \displaystyle M^{K'}_{\rm ZZ1}\left(
\begin{array}{c}
A^{K'}
\\
B^{K'}_+
\\
B^{K'}_-
\end{array}
\right)
= 0, \\
& \displaystyle
M^{K'}_{\rm ZZ1}\equiv
\left(
\begin{array}{ccc}
\alpha_2 \phi^L_{\nu} & 
\beta^+_4 \,\phi^R_{\nu_+ +1} &
\beta^-_4 \,\phi^R_{\nu_- +1} 
\\
- \alpha_1 \phi^L_{\nu-1} & 
\beta^+_3 \,\phi^R_{\nu_+} &
\beta^-_3 \,\phi^R_{\nu_-} 
\\
0 & 
\beta^+_1 \,\phi^R_{\nu_+ -1} &
\beta^-_1 \,\phi^R_{\nu_- -1} 
\end{array}
\right).
\label{eq_eigen_zz1_K'}
\end{eqnarray}

\subsection{Zigzag boundary, ZZ2}

The boundary ZZ2 is another zigzag boundary where
the front-most line of bilayer part is formed by $B_2$ sites.
The boundary conditions are\cite{Nakanishi_et_al_2010a}
\begin{eqnarray}
&& F^v_{A1}(0,y) =  F^v_A(0,y), \nonumber\\
&& F^v_{B1}(0,y) =  F^v_B(0,y), \\
&& F^v_{A2}(0,y) =  0, \nonumber
\label{eq_bc_zz2}
\end{eqnarray}
where only the third condition is different
from Eq.\ (\ref{eq_bc_zz1}).
% There is no restriction for $F^v_{B2}(0,y)$ instead.
Similarly to ZZ1, we obtain the matrix
for the $K$ and $K'$ points
\begin{eqnarray}
& \displaystyle
M^{K}_{\rm ZZ2}\equiv
\left(
\begin{array}{ccc}
\alpha_1 \phi^L_{\nu-1} & 
\beta^+_1 \,\phi^R_{\nu_+ -1} &
\beta^-_1 \,\phi^R_{\nu_- -1} 
\\
- \alpha_2 \phi^L_{\nu} & 
\beta^+_2 \,\phi^R_{\nu_+} &
\beta^-_2 \,\phi^R_{\nu_-} 
\\
0 & 
\beta^+_3 \,\phi^R_{\nu_+} &
\beta^-_3 \,\phi^R_{\nu_-} 
\end{array}
\right), \\
\label{eq_eigen_zz2_K}
& \displaystyle
M^{K'}_{\rm ZZ2}\equiv
\left(
\begin{array}{ccc}
\alpha_2 \phi^L_{\nu} & 
\beta^+_4 \,\phi^R_{\nu_+ +1} &
\beta^-_4 \,\phi^R_{\nu_- +1} 
\\
- \alpha_1 \phi^L_{\nu-1} & 
\beta^+_3 \,\phi^R_{\nu_+} &
\beta^-_3 \,\phi^R_{\nu_-} 
\\
0 & 
\beta^+_2 \,\phi^R_{\nu_+} &
\beta^-_2 \,\phi^R_{\nu_-} 
\end{array}
\right). 
\label{eq_eigen_zz2_K'}
\end{eqnarray}
\par
%%%%%%%%%%%%%%%%%%%%%%%%%%%%%%%%%%%%%%%%%%%%%%%%%%%%%%%%%%%%%%%%%%%%%%%%%%%%%%%
%
%%%%%%%%%%%%%%%%%%%%%%%%%%%%%%%%%%%%%%%%%%%%%%%%%%%%%%%%%%%%%%%%%%%%%%%%%%%%%%%
\subsection{Armchair boundary (AC)}
%%%%%%%%%%%%%%%%%%%%%%%%%%%%%%%%%%%%%%%%%%%%%%%%%%%%%%%%%%%%%%%%%%%%%%%%%%%%%%%
%
%%%%%%%%%%%%%%%%%%%%%%%%%%%%%%%%%%%%%%%%%%%%%%%%%%%%%%%%%%%%%%%%%%%%%%%%%%%%%%%
The boundary conditions for the armchair boundary AC are\cite{Nakanishi_et_al_2010a}
\begin{equation}
\begin{array}{l}
F^v_{A1}(0,y) =  F^v_A(0,y), \\
F^v_{B1}(0,y) =  F^v_B(0,y), \\
F^K_{A2}(0,y) - F^{K'}_{A2}(0,y) =  0, \\
F^K_{B2}(0,y) + F^{K'}_{B2}(0,y) =  0,
\end{array}
\label{eq_bc_arm}
\end{equation}
where the third and fourth conditions mix the wavefunctions at the $K$ and $K'$ points.
They are rewritten as 
\begin{eqnarray}
M_{\rm AC}\left(
\begin{array}{c}
A^K \\ B^K_+ \\ B^K_- \\ 
A^{K'} \\ B^{K'}_+ \\ B^{K'}_-
\end{array}
\right)
= 0 ,
\end{eqnarray}
with
\begin{eqnarray}
\displaystyle M_{\rm AC} &=& \left(\begin{array}{cc} M^K_{\rm ZZ1} & M^{KK'} \\ M^{K'K} & M^{K'}_{\rm ZZ2} \end{array} \right) , \\
\displaystyle M^{KK'} &=& \left( \begin{array}{ccc} 0 & 0 & 0 \\ 0 & 0 & 0 \\ 0 & - \beta_1^+ \phi_{\nu_+-1}^R & - \beta_1^- \phi_{\nu_--1}^R \end{array} \right) , \\
\displaystyle M^{K'K} &=& \left( \begin{array}{ccc} 0 & 0 & 0 \\ 0 & 0 & 0 \\ 0 & \beta_3^+ \phi_{\nu_+}^R & \beta_3^- \phi_{\nu_-}^R \end{array} \right) .
\end{eqnarray}
\par
%%%%%%%%%%%%%%%%%%%%%%%%%%%%%%%%%%%%%%%%%%%%%%%%%%%%%%%%%%%%%%%%%%%%%%%%%%%%%%%
%
%%%%%%%%%%%%%%%%%%%%%%%%%%%%%%%%%%%%%%%%%%%%%%%%%%%%%%%%%%%%%%%%%%%%%%%%%%%%%%%
\subsection{Interface Landau levels}
\label{sec_flat}
%%%%%%%%%%%%%%%%%%%%%%%%%%%%%%%%%%%%%%%%%%%%%%%%%%%%%%%%%%%%%%%%%%%%%%%%%%%%%%%
%
%%%%%%%%%%%%%%%%%%%%%%%%%%%%%%%%%%%%%%%%%%%%%%%%%%%%%%%%%%%%%%%%%%%%%%%%%%%%%%%
Let us consider a special state of valley $v$,
which satisfies the conditions
\begin{equation}
\begin{array}{l}
F^v_{A1}(0,y) =  F^v_A(0,y), \\
F^v_{B1}(0,y) =  F^v_B(0,y), \\
F^v_{A2}(0,y) =  0, \\
F^v_{B2}(0,y) =  0.
\end{array}
\label{eq_bc_flat}
\end{equation}
Because these include both boundary conditions 
for ZZ1 and ZZ2,
such a state must be shared by both ZZ1 and ZZ2.
Those states exist at different series of points $(\varepsilon, X)$
for $v=K$ and $K'$, denoted by $P_K$ and $P_{K'}$, respectively.
Further, the wavefunction satisfying Eq.\ (\ref{eq_bc_flat}) at 
valley $v$
also meets conditions (\ref{eq_bc_arm}) for the armchair boundary,
when the wave amplitudes of the other valley 
(opposite valley of $v$)
are all zero.
As a result, points $P_K$ and $P_{K'}$
are also shared by Landau levels in an armchair boundary.
\par
%%%%%%%%%%%%%%%%%%%%%%%%%%%%%%%%%%%%%%%%%%%%%%%%%%%%%%%%%%%%%%%%%%%%%%%%%%%%%%%
%
%%%%%%%%%%%%%%%%%%%%%%%%%%%%%%%%%%%%%%%%%%%%%%%%%%%%%%%%%%%%%%%%%%%%%%%%%%%%%%%
Using some algebra, we can show that
at the special points $P_K$ and $P_{K'}$,
the gradient of the Landau level energy in $X$ 
vanishes in any types of boundaries ZZ1, ZZ2, and AC.
We can show that the second derivative also vanishes for ZZ2.
The detailed proof is presented in Appendix \ref{app_flat}.
Accordingly the density of states diverges at the identical energies
independent of the boundary type.
Further, at those points, the wavefunctions of monolayer part and bilayer part connect smoothly on layer 1, because the amplitude on layer 2 locally vanishes and thus hardly affects the electron motion on layer 1.
As a result, the wavefunctions on the monolayer and bilayer sides 
are coupled well, and the amplitude is almost equally distributed to both sides.
\par
%%%%%%%%%%%%%%%%%%%%%%%%%%%%%%%%%%%%%%%%%%%%%%%%%%%%%%%%%%%%%%%%%%%%%%%%%%%%%%%
%
%%%%%%%%%%%%%%%%%%%%%%%%%%%%%%%%%%%%%%%%%%%%%%%%%%%%%%%%%%%%%%%%%%%%%%%%%%%%%%%
As will be demonstrated in numerical results presented in the next section, 
in ZZ1 and AC, $\varepsilon(X)$ takes a local maximum at 
each $P_K$ and $P_{K'}$ in positive energies, while 
there usually exists another point nearby 
where $\varepsilon(X)$ takes a local minimum, 
giving divergent density of states as well.
We will show that, around these points,
a crossover takes place from a monolayer edge-state mainly 
localized in monolayer,
to a bilayer edge-state mainly localized in bilayer, 
when $X$ is varied.
It is natural that slight shift in $X$ 
does not change the energy at such crossover points, 
because they are anti-crossing points between intersecting 
energy levels of monolayer and bilayer edgestates.
In ZZ2, the energy minima and maxima are degenerate
corresponding to vanishing second derivative,
and thus $\varepsilon(X)$ is even smoother 
and the divergence in the density of states is stronger than ZZ1 and AC.
These nearly flat-band regions around extrema of $\varepsilon(X)$ 
can be referred to as the interface Landau levels.

\par
%%%%%%%%%%%%%%%%%%%%%%%%%%%%%%%%%%%%%%%%%%%%%%%%%%%%%%%%%%%%%%%%%%%%%%%%%%%%%%%
%
%%%%%%%%%%%%%%%%%%%%%%%%%%%%%%%%%%%%%%%%%%%%%%%%%%%%%%%%%%%%%%%%%%%%%%%%%%%%%%%
\subsection{Zero energy levels}
\label{sec_zero}
%%%%%%%%%%%%%%%%%%%%%%%%%%%%%%%%%%%%%%%%%%%%%%%%%%%%%%%%%%%%%%%%%%%%%%%%%%%%%%%
%
%%%%%%%%%%%%%%%%%%%%%%%%%%%%%%%%%%%%%%%%%%%%%%%%%%%%%%%%%%%%%%%%%%%%%%%%%%%%%%%
The energy spectrum of a monolayer-bilayer junction
approaches that of bulk monolayer and bilayer graphenes
in the limit of $X \to +\infty$ and $-\infty$, respectively,
because the wave function, centered at $x = X$, 
mostly resides in the bulk region far from the boundary.
On the other hand,
the zero-energy level is special in that it is contributed
not only by the bulk Landau levels, 
but also by the zero-energy edge states, 
which are localized near the boundary region
on the terminated layer of bilayer graphene. \cite{Nakanishi_et_al_2010a}
We can analytically obtain 
the energies and wavefunctions of zero energy Landau levels
using the above formulation, as demonstrated
in Appendix \ref{app_zero} for ZZ2 boundary.
Table \ref{tbl_zero} summarizes the degeneracy of zero energy levels
in the limit of $X \to \pm\infty$
for each boundary type,
where $+1$ represents the additional degeneracy due to the edge states.
In ZZ1 and ZZ2, the edge state appears either of $X=\pm\infty$
depending on valleys, while it is absent in AC. \cite{Nakanishi_et_al_2010a}

\begin{table}
\begin{tabular}{lcc}
& (a) ZZ1 &
\\
$X$  & $-\infty$ & $+\infty$ 
\\ \hline
$K$   & 1 & 2+1
\\ \hline
$K'$  & 1+1 & 2 
\end{tabular}
\quad
\begin{tabular}{lcc}
& (b) ZZ2 &
\\
$X$  & $-\infty$ & $+\infty$ 
\\ \hline
$K$   & 1+1 & 2 
\\ \hline
$K'$  & 1 & 2+1 
 \end{tabular}
\quad
\begin{tabular}{lcc}
& (c) AC &
\\
$X$  & $-\infty$ & $+\infty$ 
\\ \hline
$K$   & 1 & 2 
\\ \hline
$K'$  & 1 & 2 
 \end{tabular}
\caption{Number of zero-energy Landau levels per spin 
in the limit of $X = \pm\infty$,
for (a) zigzag-1, (b) zigzag-2, and (c) armchair boundaries.
$+1$ represents extra degeneracy due to the zero-energy edge mode.
}
\label{tbl_zero}
\end{table}

\begin{figure*}
\includegraphics[width=120mm]{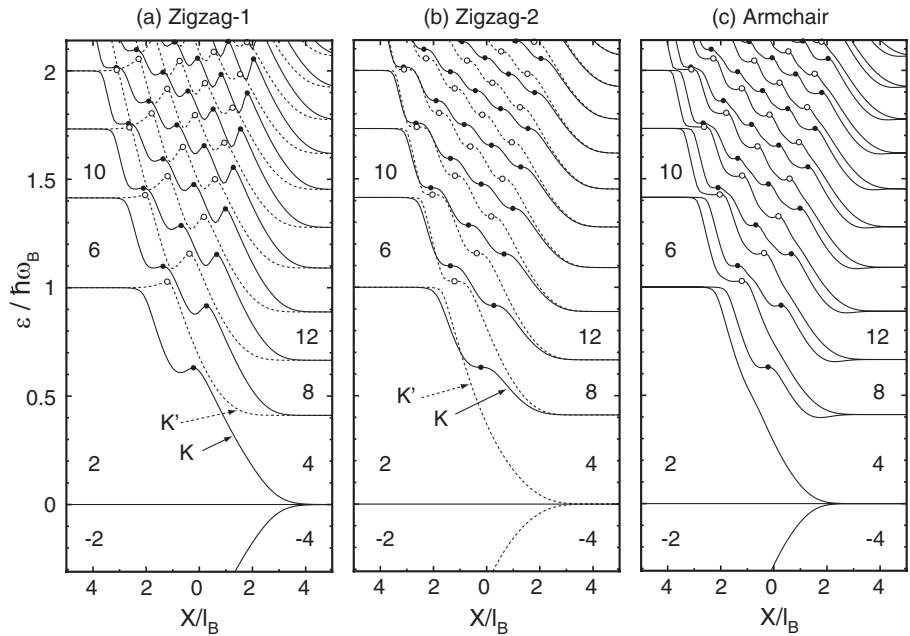}
\caption{
Energy spectrum as a function of $X$ in boundary 
(a) ZZ1, (b) ZZ2, and (c) AC,
at magnetic field $\hbar\omega_B = \gamma_1/3$ $(B\sim 10\textrm{T})$.
Black and white circles represent $P_K$ and $P_{K'}$, respectively.
Numbers between the levels
indicate bulk filling factor in limits $X = \pm\infty$. 
}
\label{fig_band}
\end{figure*}

\subsection{Local density of states}
\label{sec_ldos}

In monolayer graphene, the amplitude of the wavefunctions at $A$ and $B$
sites are written in terms of effective-mass envelope functions as \cite{Ando_2005a}
\begin{equation}
\begin{array}{l}
\psi_{A}({\bf R}) = e^{ i {\bf K}\cdot{\bf R}} F_{A}^{K}({\bf R}) +
 e^{ i \eta} e^{ i {\bf K}'\cdot{\bf R}} F_{A}^{K'}({\bf R}) , \\
\noalign{\vspace{0.1250cm}}
\psi_{B}({\bf R}) = - \omega e^{ i \eta} e^{i {\bf K}\cdot{\bf R}} F_{B}^{K}({\bf R}) + e^{ i {\bf K}'\cdot{\bf R}} F_{B}^{K'}({\bf R}) , 
\end{array}
\label{eq_wave_atom_mono}
\end{equation}
where $\eta$ is the angle between the $x$ axis and zigzag direction of honeycomb lattice and $\omega=e^{2\pi i /3}$.
In bilayer graphene, the amplitude can be written as \cite{Nakanishi_et_al_2010a}
\begin{eqnarray}
&&\psi_{A1}({\bf R})  \! = e^{i {\bf K}\cdot{\bf R}} F_{A1}^{K}({\bf R})
 + e^{i \eta} e^{i {\bf K}'\cdot{\bf R}} F_{A1}^{K'}({\bf R}) , 
\nonumber\\
&&\psi_{B1}({\bf R})  \! = - \omega e^{i \eta} e^{i {\bf K}\cdot{\bf R}}
 F_{B1}^{K}({\bf R}) + e^{i {\bf K}'\cdot{\bf R}} F_{B1}^{K'}({\bf R}) , 
\nonumber\\
&&\psi_{A2}({\bf R})  \! = - \omega e^{i \eta}e^{i {\bf K}\cdot{\bf R}}
 F_{A2}^{K}({\bf R}) + e^{i {\bf K}'\cdot{\bf R}} F_{A2}^{K'}({\bf R}) , 
\nonumber\\
&&\psi_{B2}({\bf R})  \! = \omega^{-1} e^{2i \eta} e^{i {\bf K}\cdot{\bf R}} F_{B2}^{K}({\bf R}) + e^{-i \eta}e^{i {\bf K}'\cdot{\bf R}} F_{B2}^{K'}({\bf R}) .
\nonumber\\
\label{eq_wave_atom_bi}
\end{eqnarray}

\begin{figure}
\includegraphics[width=80mm]{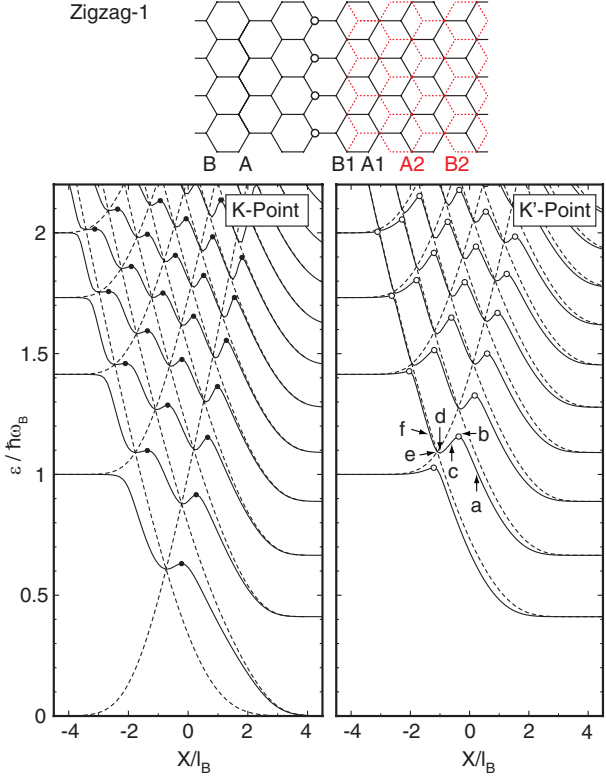}
\caption{(Above) Monolayer-bilayer junction of type ZZ1,
separated by infinite potential on white circles
into independent monolayer and bilayer graphenes.
(Below) Energy spectrum of junction ZZ1 (solid) 
and separated system (dashed),
in magnetic field $\hbar\omega_B = \gamma_1/3$.
Left and right panels show the spectra of the $K$ and $K'$ points,
respectively,
and the black and white circles are the interface Landau levels.
}
\label{fig_crossing_zig1}
\end{figure}

\begin{figure}
\includegraphics[width=80mm]{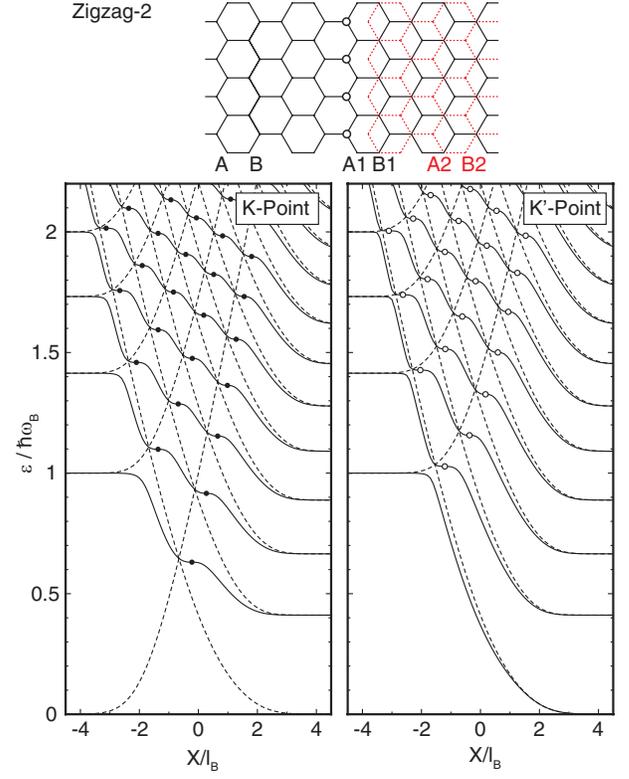}
\caption{Plots similar to Fig.\ \ref{fig_crossing_zig1}
for ZZ2.
}
\label{fig_crossing_zig2}
\end{figure}

In a tight-binding model, 
the local density of state (LDOS) on site $A$ located at
the position $\Vec{R}$ is defined by
\begin{equation}
 \rho_A(\varepsilon; \Vec{R}) = 
\sum_\alpha 
\delta(\varepsilon-\varepsilon^{(\alpha)})
|\psi^{(\alpha)}_A(\Vec{R})|^2,
\label{eq_ldos}
\end{equation}
where $\varepsilon^{(\alpha)}$ and $\psi^{(\alpha)}$ are the energy and the wavefunction of eigenstate $\alpha$.
Similar expressions can be written down for other sites $B$, $A1$, etc.
When the wave amplitudes at the $K$ and $K'$ points coexist 
in a single eigenstate,
LDOS has a Kekul\'{e} pattern
due to the interference between the factors
$e^{ i {\bf K}\cdot{\bf R}}$ 
and $e^{ i {\bf K'}\cdot{\bf R}}$.
\cite{Nakanishi_and_Ando_2008a}
In the present case, this is expected to appear
in the armchair boundary which mixes the $K$ and $K'$ valleys,
while absent in ZZ1 or ZZ2, where every eigenstate
is a single-valley state.
We also define the spatially averaged LDOS for site $A$ as
\begin{equation}
 \rho^{\rm av}_A(\varepsilon; \Vec{R}) = 
\sum_\alpha 
\sum_{v=K,K'}
\delta(\varepsilon-\varepsilon^{(\alpha)}) 
|(F^{(\alpha)})^v_{A}(\Vec{R})|^2.
\label{eq_ldos_av}
\end{equation}
This is an average of the original LDOS in Eq.\ (\ref{eq_ldos})
over several unit cells in the region smaller than typical length scales of the envelope function.
\par
%%%%%%%%%%%%%%%%%%%%%%%%%%%%%%%%%%%%%%%%%%%%%%%%%%%%%%%%%%%%%%%%%%%%%%%%%%%%%%%
%
%%%%%%%%%%%%%%%%%%%%%%%%%%%%%%%%%%%%%%%%%%%%%%%%%%%%%%%%%%%%%%%%%%%%%%%%%%%%%%%
\section{Numerical results}
\label{sec_num}
%%%%%%%%%%%%%%%%%%%%%%%%%%%%%%%%%%%%%%%%%%%%%%%%%%%%%%%%%%%%%%%%%%%%%%%%%%%%%%%
%
%%%%%%%%%%%%%%%%%%%%%%%%%%%%%%%%%%%%%%%%%%%%%%%%%%%%%%%%%%%%%%%%%%%%%%%%%%%%%%%
Figure \ref{fig_band} shows
the energy spectra against $X$,
numerically calculated for the junctions of
ZZ1, ZZ2, and AC boundaries 
at magnetic field of $\hbar\omega_B = \gamma_1/3$
($B\sim 10$T).
Landau levels approach those of bulk monolayer and of bilayer
in the limit $X \to \infty$ and $-\infty$, respectively.
In the boundary region, the valley-degenerate levels split
and connect to different levels in the opposite side.
The black and white circles represent the interface Landau levels
$P_K$ and $P_{K'}$, respectively,
which are independent of boundary type and corresponds to local band maxima.
In accordance with the argument in Sec.\ \ref{sec_flat},
we actually see that energy levels pass through those points
in all three cases
and the gradient vanishes there.
In ZZ1 and AC, the band minima are also present near the maxima at
$P_K$ and $P_K'$,
while in ZZ2 the minima and the maxima merge into inflection points 
as the second derivative vanishes.
\par
%%%%%%%%%%%%%%%%%%%%%%%%%%%%%%%%%%%%%%%%%%%%%%%%%%%%%%%%%%%%%%%%%%%%%%%%%%%%%%%
%
%%%%%%%%%%%%%%%%%%%%%%%%%%%%%%%%%%%%%%%%%%%%%%%%%%%%%%%%%%%%%%%%%%%%%%%%%%%%%%%
The oscillatory band structures appearing in the boundary region
can be understood in relation to
terminated monolayer and bilayer graphenes. 
Let us take ZZ1 boundary, and
consider a system with infinite on-site potential 
added on an array of $B$ sites near the boundary,
as illustrated as white circles in the top panel of
Fig.\ \ref{fig_crossing_zig1}.
The system is then separated into monolayer terminated with Klein's edge
and bilayer terminated with zigzag edge.
In the effective mass approximation, this is equivalent to
the boundary condition $F^v_B = 0$ for monolayer 
and $F^v_{B1} = F^v_{B2} = 0$ for bilayer.
Note that in the effective-mass approximation, shifting of on-site 
potential position by the order of the lattice constant 
does not make a difference in the result.

Lower panels of Fig.\ \ref{fig_crossing_zig1} show the energy spectrum 
of ZZ1 (solid lines)
and that of the terminated system (dashed lines),
for each of $K$ and $K'$.
In the terminated system, the independent
Landau levels of monolayer and bilayer
sharply go up as $X$
goes over the boundary.\cite{Brey_and_Fertig_2006b,Abanin_et_al_2006a}
Apparently, the spectrum of ZZ1 
resembles that of the terminated system, with an energy gap opened
at every crossing point.
%by mixing between monolayer and bilayer states.
The resemblance of the two different spectra
may be attributed to following reasons.
In the monolayer-bilayer junction, 
when a low-energy electron travels from the monolayer 
to the bilayer,
it feels as if $B$ sites suddenly disappear at the boundary, 
because in bilayer, $B_1$ is coupled with $A_2$
to make high-energy states away from $\varepsilon = 0$.
This effect can be roughly modeled by on-site potential
at $B$ sites at the boundary.
For an electron coming from bilayer side, on the other hand, 
$B_1$ site which was absent in the low-energy spectrum
suddenly resumes at the beginning of the monolayer region,
while $A_1$ just smoothly connects to $A$.
This should roughly correspond to 
some condition for $B_1$ sites, with $A_1$ left intact.
Energy gaps opening at crossing points are due to finite hybridization
between monolayer and bilayer states.

As another remark,
we observe that energy levels of ZZ1 pass through
every crossing point of terminated bilayer 
and monolayer levels. This occurs when an eigenfunction of ZZ1
happens to have a node on the on-site potential sites,
because such a state is also an eigenstate when on-site potential is present.
Therefore the wavefunction of ZZ1 becomes 
identical with that of the terminated system at each crossing point.

\begin{figure*}
\includegraphics[width=180mm]{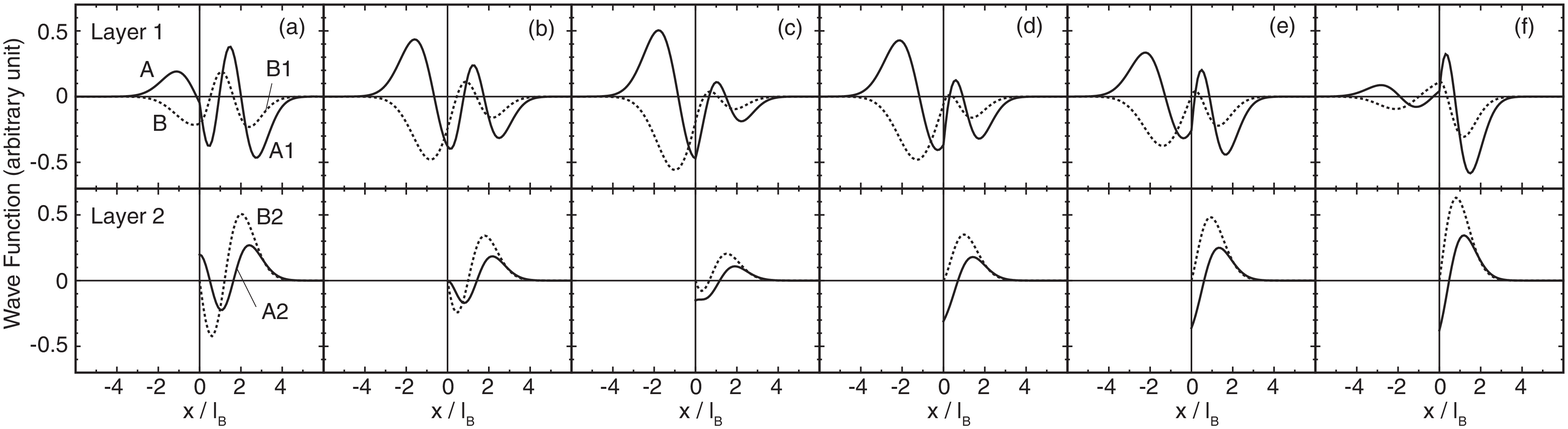}
\caption{
Wavefunctions near the interface Landau level of the $K'$ point in ZZ1 at $\hbar\omega_B = \gamma_1/3$.
(a) to (f) correspond to the points in the energy spectrum in Fig.\ \ref{fig_crossing_zig1}.
}
\label{fig_wav}
\end{figure*}

\begin{figure*}
\includegraphics[width=170mm]{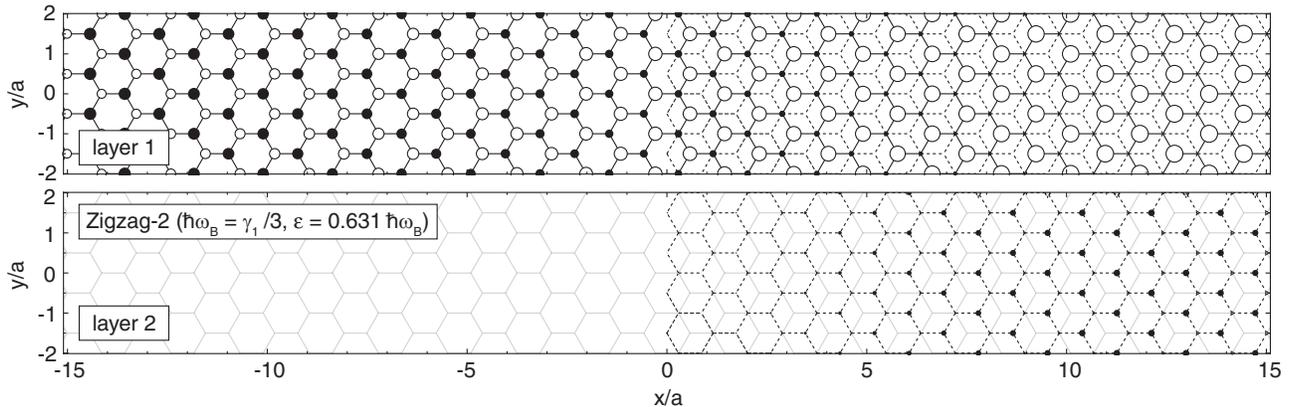}
\caption{
Local density of states
of boundary ZZ2 with $\hbar\omega_B = \gamma_1/3$,
at the energy $\varepsilon = 0.631\hbar\omega_B$
near the lowest interface Landau level of $K$.
The areas of circles in upper and lower panels
represent the amplitude of LDOS 
at each atom on the layer 1 and 2, respectively,
while open and filled circles represent $A$ and $B$ sublattices, respectively.
$l_B/a$ is about $30$ at this magnetic field.
}
\label{fig_atom_zig2}
\end{figure*}

Similar analysis is also available in boundary ZZ2.
Figure \ref{fig_crossing_zig2} compares the energy spectrum 
of ZZ2, and that of separated system
with on-site potential on $B$ sites illustrated in the top panel.
The boundary condition becomes $F^v_B = 0$ for monolayer 
and $F^v_{B1} = F^v_{A2} = 0$ for bilayer.
Since the low-energy spectrum of the bilayer is dominated 
by $A1$ and $B2$, the second condition $F^v_{A2} = 0$ should give a 
weaker effect compared to $F^v_{B2} = 0$ in ZZ1
and thus leads to better coupling between the monolayer and bilayer
region.
In Fig.\ \ref{fig_crossing_zig2}, indeed, the mixing
between terminated levels looks stronger than in ZZ1, 
resulting in the monotonic dependence 
rather than non-monotonic behavior in ZZ1.

% wavefunction near interface LL

Figure \ref{fig_wav} plots the wavefunctions
near an interface Landau levels
associated with the $K'$ point in ZZ1, where (a) to (f) correspond to
the points in the energy spectrum in Fig.\ \ref{fig_crossing_zig1}.
The point (b) is exactly at a local maximum $P_{K'}$.
There, the wavefunction of layer 1 smoothly connects at the boundary
as argued in Sec.\ \ref{sec_flat},
while generally not in other cases.
The point (e) is exactly at the crossing point
of terminated levels.
There, the wave function indeed has a node at the interface for the $B$ and $B1$ components and thus can be an eigenstate of the separate monolayer and bilayer.
At the local band minimum (d), the wave function does not have special features in contrast to $P_{K'}$.

In the energy spectrum,
the region between (a) and (b) and between (e) and (f) have slope close to that of the terminated bilayer, while the region between (b) and (d) has slope close to that of the terminated monolayer.
Correspondingly, 
the wavefunctions of (a) and (f) has significant amplitudes 
in bilayer side,
while (c) has great amplitude in monolayer side.

\begin{figure*}
\includegraphics[width=170mm]{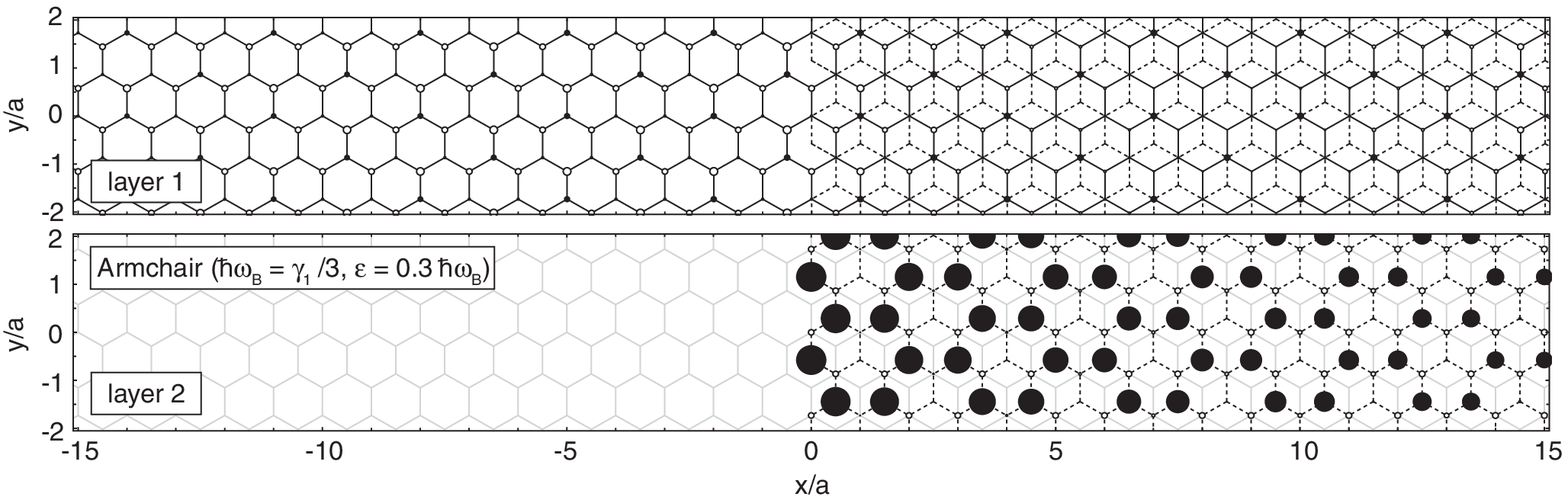}
\caption{
Plot similar to Fig.\ \ref{fig_atom_zig2},
showing the local density of states 
of the armchair boundary with $\hbar\omega_B = \gamma_1/3$
at the energy $\varepsilon = 0.3\hbar\omega_B$.
}
\label{fig_atom_arm}
\end{figure*}

\begin{figure*}
\includegraphics[width=170mm]{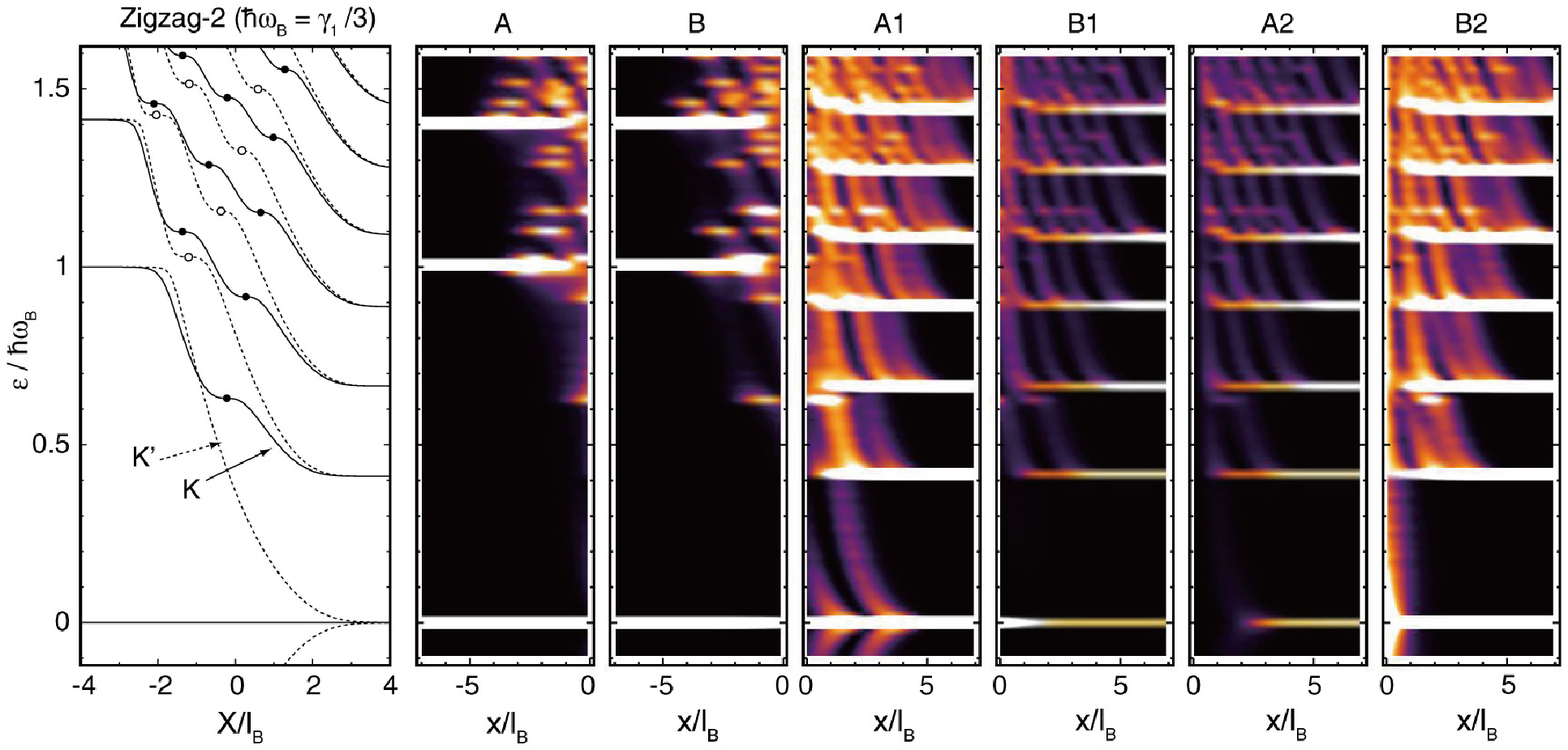}
\caption{
(Right) Averaged local density of states at different atomic sites
and (left) corresponding energy spectrum
in boundary ZZ2
at magnetic field $\hbar\omega_B = \gamma_1/3$ $(B\sim 10\textrm{T})$.
The width of each bin in energy is taken as 
$0.02\hbar\omega_B$.
}
\label{fig_ldos}
\end{figure*}

% Atomic LDOS

Figure \ref{fig_atom_zig2} illustrates
the typical atomic-scale LDOS of Eq.\ (\ref{eq_ldos})
at interface Landau levels.
We here take the boundary ZZ2 at the energy of the lowest interface Landau level
near $\varepsilon = 0.631\hbar\omega_B$ in Fig.\ \ref{fig_band}(b).
The ratio of the magnetic length to the lattice constant, $l_B/a$,
is about $30$ at this magnetic field.
The areas of circles in upper and lower panels
represent the relative amplitude of LDOS 
at each atom on layers 1 and 2, respectively,
while open and filled circles represent the A and B sublattices.
The result mainly reflects the wavefunction of interface Landau level, 
since the flat band gives 
a dominant contribution to LDOS.
We see that the wave amplitude on layer 1
connects smoothly at the boundary region,
as the amplitude of layer 2 is almost absent there.

For comparison, we show the similar plot of LDOS
of the armchair boundary at a different energy 
$\varepsilon = 0.3\hbar\omega_B$ in Fig.\ \ref{fig_atom_arm}.
In accordance with the previous argument, 
the plot clearly exhibits the Kekul\'{e} pattern unlike in ZZ2.
Note that, even in the armchair boundary, the Kekul\'{e} pattern
disappears when the energy comes to an interface Landau level,
because the eigenfunction becomes a single-valley state there.

% Averaged LDOS

Figure \ref{fig_ldos} shows
the averaged local density of states (LDOS)
defined in Eq.\ (\ref{eq_ldos_av}), for ZZ2 boundary.
The vertical scale is shared with the corresponding
energy spectrum at left.
We observe series of peaks corresponding to the 
interface Landau levels of $P_K$ and $P_{K'}$
owing to the large LDOS due to the flat band,
and its spatial distribution is
characterized by node pattern of the corresponding wavefunction.
While not shown, the peak patterns are quite 
similar among ZZ1, ZZ2, and AC, 
since every interface Landau level appears at the
identical energy with the identical effective-mass wavefunction.
In ZZ1 and AC, the band minima appearing near $P_K$ and $P_{K'}$ also contribute to the LDOS divergence and the peak structure is a little blurred.

Near the interface Landau levels,
LDOS has a considerable amplitude in monolayer region,
while otherwise it is localized mostly in the bilayer region.
This is because the monolayer and bilayer states are well hybridized
near the interface Landau levels,
while in other regions where the band lines are downslope, the states mainly originate from bilayer, as argued above.

\section{Conclusion}
\label{sec_conc}

We have studied electronic structures of 
monolayer-bilayer graphene junctions in magnetic fields.
The energy spectrum near the boundary region
is characterized by the interface Landau levels
where the band energy is locally constant,
which arise from hybridization of Landau levels of terminated monolayer
and bilayer graphenes.
The energies of interface Landau levels
are insensitive to the way the second layer is terminated,
suggesting that they would be robust 
even in a disordered junction containing
a random atomic configuration at the boundary.
Interface Landau levels give a characteristic peak pattern 
to LDOS, which may be observed by scanning spectroscopic measurement. 
\cite{Kobayashi_et_al_2005a,Niimi_et_al_2006a,Matsui_et_al_2005a,Niimi_et_al_2006b}

%%%%%%%%%%%%%%%%%%%%%%%%%%%%%%%%%%%%%%%%%%%%%%%%%%%%%%%%%%%%%%%%%%%%%%%%%%%%%%%
%
%%%%%%%%%%%%%%%%%%%%%%%%%%%%%%%%%%%%%%%%%%%%%%%%%%%%%%%%%%%%%%%%%%%%%%%%%%%%%%%
\acknowledgments
%%%%%%%%%%%%%%%%%%%%%%%%%%%%%%%%%%%%%%%%%%%%%%%%%%%%%%%%%%%%%%%%%%%%%%%%%%%%%%%
%
%%%%%%%%%%%%%%%%%%%%%%%%%%%%%%%%%%%%%%%%%%%%%%%%%%%%%%%%%%%%%%%%%%%%%%%%%%%%%%%
This work was supported in part by Grant-in-Aid for Scientific
Research on Priority Area ``Carbon Nanotube Nanoelectronics,''
by Grant-in-Aid for Scientific Research, by
Global Center of Excellence Program at Tokyo Tech ``Nanoscience
and Quantum Physics'' from the Ministry of Education,
Culture, Sports, Science and Technology, Japan,
and by JST-EPSRC Japan-UK Cooperative
Programme under Grant No.\ EP/H025804/1.

\appendix

\section{Special flat-band points}
\label{app_flat}

Here we prove that, 
at a special point $(\varepsilon,X)$
where a wavefunction satisfies Eq.\ (\ref{eq_bc_flat}),
the derivative of Landau level in $X$ vanishes 
in ZZ1, ZZ2, and AC, and
the second derivative also vanishes 
in ZZ2.
The condition Eq.\ (\ref{eq_bc_flat})
is alternatively written as
\begin{equation}
\det M^{v}_{\rm ZZ1}
=
\det M^{v}_{\rm ZZ2} = 0.
\label{eq_bc_flat2}
\end{equation}
In the following, we will show that 
Eq. (\ref{eq_bc_flat2}) leads to
\begin{eqnarray}
&& \frac{\partial}{\partial X}\det M^{v}_{\rm ZZ1}
= \frac{\partial}{\partial X}\det M_{\rm AC}
= 0
\label{eq_bc_diff1}
\\
&& 
\frac{\partial}{\partial X}\det M^{v}_{\rm ZZ2} = \frac{\partial^2}{\partial X^2}\det M^{v}_{\rm ZZ2} = 0,
\label{eq_bc_diff2}
\end{eqnarray}
which immediately proves the statements above.

In the matrices $M^{v}_{\rm ZZ1}$ and $M^{v}_{\rm ZZ2}$,
we can eliminate $\phi^{L,R}_{\nu-1}$
by replacing them with
$\phi^{L,R}_{\nu}$ and $\phi^{L,R}_{\nu+1}$,
using the recursion formula of Weber's function,
\begin{eqnarray}
 D_{\nu+1} - z D_\nu + \nu D_{\nu-1} = 0.
\label{eq_recur}
\end{eqnarray}
Eq.\ (\ref{eq_bc_flat2}) at $v=K$
can then be transformed as
\begin{eqnarray}
&& \sum_{\mu=\pm} \mu (\varepsilon^2 - \nu_{\bar{\mu}}) 
\phi^R_{\nu_{\bar{\mu}}}
(\phi^L_{\nu} \phi^R_{\nu_{\mu}+1}
+ \phi^L_{\nu+1} \phi^R_{\nu_{\mu}}
) = 0,
\label{eq_app1}
\\
&&
\sum_{\mu=\pm} \mu (\varepsilon^2 - \nu_{\bar{\mu}}) 
\phi^R_{\nu_{\bar{\mu}+1}}
(\phi^L_{\nu} \phi^R_{\nu_{\mu}+1}
+ \phi^L_{\nu+1} \phi^R_{\nu_{\mu}}
) = 0, \qquad
\label{eq_app2}
\end{eqnarray}
with $\bar{\mu} = -\mu$, leading to
\begin{eqnarray}
&& 
\phi^L_{\nu} \phi^R_{\nu_{\mu}+1}
+ \phi^L_{\nu+1} \phi^R_{\nu_{\mu}} = 0,
\label{eq_app4}
\end{eqnarray}
for both of $\mu=\pm$.

The derivative of the matrix determinant in $X$ can be evaluated using 
Eq.\ (\ref{eq_deriv}).
For $M^K_{\rm ZZ1}$, we obtain
\begin{eqnarray}
&\displaystyle \frac{\partial \det M^K_{\rm ZZ1}}{\partial X} \! \propto \!\!
\sum_{\mu=\pm} \mu (\varepsilon^2 \!-\! \nu_{\bar{\mu}}) 
\phi^{R\prime}_{\nu_{\bar{\mu}+1}}
(\phi^L_{\nu} \phi^R_{\nu_{\mu}+1}
\!+\! \phi^L_{\nu+1} \phi^R_{\nu_{\mu}}
) & \nonumber \\
&- \phi^L_{\nu} \sum_{\mu=\pm} \mu 
(\varepsilon^2 - \nu_{\mu}) 
(\varepsilon^2 - \nu_{\bar{\mu}}) 
\phi^R_{\nu_{\mu}}\phi^R_{\nu_{\bar{\mu}+1}}, &
\label{eq_det_deriv1}
\end{eqnarray}
where $'$ represents the derivative in $X$.
At the points satisfying  Eq.\ (\ref{eq_bc_flat2}),
the first term becomes zero because of Eq.\ (\ref{eq_app4}).
The second term is transformed with Eq.\ (\ref{eq_app4}) as
\begin{eqnarray}
&&
 \phi^L_{\nu+1} \sum_{\mu=\pm} \mu 
(\varepsilon^2 - \nu_{\mu}) 
(\varepsilon^2 - \nu_{\bar{\mu}}) 
\phi^R_{\nu_{\mu}}\phi^R_{\nu_{\bar{\mu}}},
\label{eq_antisym}
\end{eqnarray}
which vanishes since the argument inside the summation
is anti-symmetric in $\mu$.

For $M^K_{\rm ZZ2}$, we have
\begin{equation}
\frac{\partial \det M^K_{\rm ZZ2}}{\partial X} \propto \! \sum_{\mu=\pm} \mu (\varepsilon^2 \!-\! \nu_{\bar{\mu}}) 
\phi^{R\prime}_{\nu_{\bar{\mu}}}
(\phi^L_{\nu} \phi^R_{\nu_{\mu}+1} \!+\! \phi^L_{\nu+1} \phi^R_{\nu_{\mu}} ),
\label{eq_det_deriv2}
\end{equation}
which similarly vanishes under the condition Eq.\ (\ref{eq_app4}).
Eq.\ (\ref{eq_det_deriv2}) is even differentiated as
\begin{eqnarray}
& \displaystyle \frac{\partial^2 \det M^K_{\rm ZZ2}}{\partial X^2}
\propto \! \sum_{\mu=\pm} \mu (\varepsilon^2 \!-\! \nu_{\bar{\mu}}) 
\phi^{R\prime\prime}_{\nu_{\bar{\mu}}}
(\phi^L_{\nu} \phi^R_{\nu_{\mu}+1}
\!+\! \phi^L_{\nu+1} \phi^R_{\nu_{\mu}}
) & \nonumber\\
& - \phi^L_{\nu} \sum_{\mu=\pm} \mu 
(\varepsilon^2 - \nu_{\mu}) 
(\varepsilon^2 - \nu_{\bar{\mu}}) 
\phi^R_{\nu_{\mu}} \phi^{R\prime}_{\nu_{\bar{\mu}}} . &
\label{eq_det_deriv2-2}
\end{eqnarray}
The first term becomes zero again under Eq.\ (\ref{eq_app4}).
The factor $(\phi^R_{\nu_{\bar{\mu}}})'$ in the second term
can be written in terms
$\phi^R_{\nu_{\bar{\mu}}}$ and $\phi^R_{\nu_{\bar{\mu}}+1}$
using Eq. (\ref{eq_deriv}).
Then it is shown to vanish
by similar transformation to Eq.\ (\ref{eq_antisym}).

The determinant of $M_{\rm AC}$
can be written in terms of those of ZZ1 and ZZ2 as,
\begin{equation}
 \det M_{\rm AC} = 
\det M^{K}_{\rm ZZ1}\det M^{K'}_{\rm ZZ2}
-
\det M^{K}_{\rm ZZ2}\det M^{K'}_{\rm ZZ1}.
\label{eq_det_arm}
\end{equation}
Under the condition 
$\det M^K_{\rm ZZ1}=\det M^K_{\rm ZZ2}=(\det M^K_{\rm ZZ1})'=(\det M^K_{\rm ZZ2})'=0$,
Eq.\ (\ref{eq_det_arm}) immediately gives $(\det M_{\rm AC})'=0$.

\section{Nearly-zero energy states}
\label{app_zero}

Let us focus on the eigenstates in the vicinity of zero energy,
taking the case of ZZ2 as an example.
We will show here that the zero-energy levels 
in monolayer-bilayer junction are contributed
not only by the Landau levels in bulk monolayer and bilayer,
but also by the zero-energy edge states localized to the boundary. 
For $K$-point, there are two independent wavefunctions 
exactly at zero energy,
\begin{eqnarray}
\Psi^{K1} =
\left\{
\begin{array}{ll}
\left(
\begin{array}{c}
F_A^K
\\
F_B^K
\end{array}
\right)
=
\left(
\begin{array}{c}
0
\\
\phi_{0} 
\end{array}
\right)
& (x<0);
\\
\left(
\begin{array}{c}
F_{A1}^K
\\
F_{B1}^K
\\
F_{A2}^K
\\
F_{B2}^K
\end{array}
\right)
=
\left(
\begin{array}{c}
0
\\
\phi_{0} / \tilde\gamma_1
\\
0
\\
-i \,\phi_{1} 
\end{array}
\right)
& (x>0),
\end{array}
\right.
\end{eqnarray}
and
\begin{eqnarray}
\Psi^{K2} =
\left\{
\begin{array}{ll}
\left(
\begin{array}{c}
F_A^K
\\
F_B^K
\end{array}
\right)
=
\left(
\begin{array}{c}
0
\\
0
\end{array}
\right)
& (x<0);
\\
\left(
\begin{array}{c}
F_{A1}^K
\\
F_{B1}^K
\\
F_{A2}^K
\\
F_{B2}^K
\end{array}
\right)
=
\left(
\begin{array}{c}
0
\\
0
\\
0
\\
\phi_{0} 
\end{array}
\right)
& (x>0),
\end{array}
\right.
\end{eqnarray}
with $\phi_{n} \equiv \phi^R_{n} = (-1)^n \phi^L_{n}$
for a nonnegative integer $n$,
and the overall normalization factor is omitted.

In $X \to \infty$, i.e., when
the center coordinate goes deep inside of the bilayer region, 
$\Psi^{K1}$ and $\Psi^{K2}$ approach the
wavefunctions of two zero-energy Landau levels of bulk bilayer graphene.
In $X \to -\infty$, on the other hand,
$\Psi^{K1}$ becomes the only zero-energy level of the bulk monolayer,
while $\Psi^{K2}$ does not have any amplitudes in the monolayer side,
but mostly concentrated on $B_2$ sites near the boundary.
$\Psi^{K2}$ at $B_2$ then approximates
\begin{eqnarray}
 \phi_{0} \propto e^{-(x-X)^2/l_B^2}
\approx {\rm const.} \times e^{- k_y x},
\end{eqnarray}
which is independent of magnetic field.
This corresponds to the zero-energy edge mode
in zero magnetic field limit. \cite{Nakanishi_et_al_2010a}

For $K'$-point, we have a single state at zero energy,
\begin{eqnarray}
\Psi^{K'1} =
\left\{
\begin{array}{ll}
\left(
\begin{array}{c}
F_A^K
\\
F_B^K
\end{array}
\right)
=
\left(
\begin{array}{c}
\phi_{0} 
\\
0
\end{array}
\right)
& (x<0);
\\
\left(
\begin{array}{c}
F_{A1}^K
\\
F_{B1}^K
\\
F_{A2}^K
\\
F_{B2}^K
\end{array}
\right)
=
\left(
\begin{array}{c}
\phi_{0} 
\\
0
\\
0
\\
0
\end{array}
\right)
& (x>0).
\end{array}
\right.
\end{eqnarray}
When $X$ moves from $-\infty$ to $+\infty$,
the wavefunction $\Psi^{K'1}$ crosses over from
the only zero-energy level in monolayer to
one of zero-energy levels in bilayer, $\nu_-=-1$.

Besides, for positive large $X$,
there exist another two levels near zero energy,
which are expressed as a hybridization of
bilayer's Landau level of $\nu_+=0$
and a zero-energy edge state.
The derivation goes as follows.
By expanding $\nu_\pm$ in Eq.\ (\ref{eq_nu_pm})
in $\varepsilon$,
the determinant of $M^{K'}_{\rm ZZ2}$ can be written
in a small $|\varepsilon|$ as,
\begin{equation}
\det M^{K'}_{\rm ZZ2}
=
- \frac{\varepsilon}{\gamma_1}\phi_{\nu_+}^R
(\phi_{\nu}^L \phi_{\nu_-}^R
+\phi_{\nu-1}^L \phi_{\nu_-+1}^R) + O(\varepsilon^3).
\label{eq_det_exp}
\end{equation}

The energies of the nearly zero-energy states in question
are given by the condition $\phi^R_{\nu_+} = 0$, 
when Eq.\ (\ref{eq_det_exp}) vanishes.
For a large $X$, the function $\phi_{\nu_+}^R$ 
can be evaluated by the asymptotic expansion 
of $D_\nu(z)$ which stands for large $|z|$:
\begin{eqnarray}
 && D_\nu(z) \sim 
\left\{
\begin{array}{l}
D^{(1)}_\nu(z) \quad (|\arg z| < 3\pi/4)
\\
D^{(1)}_\nu(z) + e^{\pm\nu \pi i} D^{(2)}_\nu(z)
\\
\qquad\qquad (\pi/4 < \pm \arg z < 5\pi/4)
\end{array}
\right. \\
 && D^{(1)}_\nu(z)  = e^{-z^2/4} z^\nu 
\sum_{k=0}^\infty
(-1)^k
\frac{\nu(\nu-1)\cdots(\nu-2k+1)}{k! 2^k z^{2k}} \nonumber \\
\noalign{\vspace{-0.3750cm}}
\\
\noalign{\vspace{-0.1250cm}}
&& D^{(2)}_\nu(z) = - \frac{\sqrt{2\pi}}{\Gamma(-\nu)}
e^{z^2/4} z^{-\nu-1}
\nonumber\\
&&\hspace{20mm}\times\sum_{k=0}^\infty 
\frac{(\nu+1)(\nu+2)\cdots(\nu+2k)}{k! 2^k z^{2k}}.
\end{eqnarray}
When $z$ is negative and $|\nu|$ is small, it approximates
\begin{eqnarray}
D_\nu(z) \approx e^{-z^2/4} + \sqrt{2\pi}\nu\, \frac{e^{z^2/4}}{z}.
\end{eqnarray}
This leads to an approximate expression $\phi_{\nu_+}^R(x=0)$ 
for positive large $X$,
\begin{eqnarray}
\phi_{\nu_+}^R 
\approx e^{-(X/l_B)^2/2} - \sqrt{2\pi}\,\nu_+ \, 
\frac{e^{(X/l_B)^2/2}}{\sqrt{2}X/l_B}.
\end{eqnarray}
$\phi_{\nu_+}^R $ becomes zero at 
$\nu_+ = (X/l_B) e^{-(X/l_B)^2}/\sqrt{\pi}$,
giving the energies of nearly-zero energy mode,
\begin{eqnarray}
&& \varepsilon^{K'2}_\pm \approx 
\pm \sqrt{\frac{X/l_B}{\sqrt{\pi}(1+\gamma_1^2)} e^{-(X/l_B)^2}},
\label{eq_nearly_zero}
\end{eqnarray}
where we used $\nu_+(\vare)\approx(1+\tilde\gamma_1^2)\vare$
for small $\vare$.
The corresponding wavefunctions in the bilayer part are written as
\begin{eqnarray}
&& \Psi^{K'2}_\pm \approx \Psi^{K'2}_{\rm bulk} \pm \Psi^{K'2}_{\rm edge}, \\
&& \Psi^{K'2}_{\rm bulk} =
\left(
\begin{array}{c}
i \phi_{1+\nu_+} 
\\
0
\\
- \phi_{0+\nu_+} /\tilde\gamma_1
\\
0
\end{array}
\right), \\
&& \Psi^{K'2}_{\rm edge} =
\left(
\begin{array}{c}
0
\\
|\tilde\varepsilon| \phi_{0+\nu_+} 
\\
0
\\
i (1+\gamma_1^2)(|\tilde\varepsilon|/\tilde\gamma_1) \phi_{-1+\nu_+} 
\end{array}
\right).
\end{eqnarray}
%with $\nu_+= (1+\gamma_1^2)\varepsilon^2$.
In $X\to\infty$, 
the energy $\varepsilon^{K'2}_\pm$
becomes exponentially small
and $\Psi^{K'2}_{\rm bulk}$ coincides with the 
zero-energy Landau level of bilayer, $\nu_+=0$.
For $\Psi^{K'2}_{\rm edge}$,
$B2$ component is nearly proportional
to $D_{-1}(z)$, and
approximates $\propto e^{-k_y x}$ near $x=0$.
This is a zero-energy edge state
localized near the boundary in the bilayer region.
\cite{Nakanishi_et_al_2010a}
Thus the states $\Psi^{K'2}_\pm$ are
described as a hybridization of the bulk bilayer Landau level
and zero-energy edge states.

%###############

\end{document}